\documentclass[aps,twocolumn,nofootinbib,preprintnumbers,citeautoscript,floatfix]{revtex4-1}
\usepackage{amsmath,amssymb,epsfig}
\usepackage{enumerate}
\usepackage{graphicx}
\usepackage{float}
\usepackage{url}
\usepackage{amsmath,amssymb,epsfig}
\usepackage{gensymb}
\begin{document}
\title{Measuring the deviation from the superposition principle in interference experiments} 
\author{G.Rengaraj$^{1}$}
\author{U.Prathwiraj$^{1}$}
\author{Surya Narayan Sahoo$^{1}$}
\author{R.Somashekhar$^{1}$}
\author{Urbasi Sinha*$^{1}$,$^{2}$}
\affiliation{$^{1}$Raman Research Institute, Sadashivanagar, Bangalore, India}
\affiliation{$^{2}$Institute for Quantum Computing, 200 University Avenue West, Waterloo, Ontario, Canada}
\email{usinha@rri.res.in}

\begin{abstract}
The Feynman Path Integral formalism has long been used for calculations of probability amplitudes. Over the last few years, it has been extensively used to theoretically demonstrate that the usual application of the superposition principle in slit based interference experiments is often incorrect. This has caveat in both optics and quantum mechanics where it is often naively assumed that the boundary condition represented by slits opened individually is same as them being opened together. The correction term comes from exotic sub leading terms in the Path Integral which can be described by what are popularly called non-classical paths. In this work, we report an experiment where we have a controllable parameter that can be varied in its contribution such that the effect due to these non-classical paths can be increased or diminished at will. Thus, the reality of these non-classical paths is brought forth in a classical experiment using microwaves, thereby proving that the boundary condition effect being investigated transcends the classical-quantum divide. We report the first measurement of a deviation (as big as $6\%$) from the superposition principle in the microwave domain using antennas as sources and detectors of the electromagnetic waves. We also show that our results can have potential applications in astronomy.
\end{abstract}

\maketitle

Young's double slit experiment plays a pivotal role in Physics especially Optics and Quantum Physics. In Classical Optics, it demonstrates the wave nature of light.  In the quantum mechanical domain when performed using single particles, it is a classic demonstration of the wave-particle duality of light and matter. Nobel laureate Richard Feynman famously said that it contains within it all the mysteries of Quantum Mechanics. Having said that, how well do we understand the various facets of the double slit experiment?\\
One of the assumptions that is commonly made in expositions of the double slit experiment is about superposition of the solutions to the wave equation with slits opened individually being the same as the solution with both slits opened at the same time. For instance, in a double slit set-up, if the solution to the wave equation with slit A open is  given by amplitude $\phi_A$ and the solution with slit B open is given by amplitude  $\phi_B$, then the solution when both slits are simultaneously open is commonly taken to be $\phi_A +\phi_B$ \cite{book1,book2,book3,bornwolf,feynman} . This is done both in the classical domain for instance superposing solutions to the Maxwell equations as well as in the quantum domain for instance superposing solutions to the Schrodinger equation. The fact is that this naive application of the superposition principle is not strictly true in both classical and quantum domains as solutions can only be superposed when they satisfy the same boundary conditions. This was quantified by \cite{PRL} where the authors theoretically quantified the deviation in terms of the normalized version of the Sorkin parameter $\kappa$ \cite{sorkin}, which turns out to be non-zero when the boundary conditions are correctly taken into account. In their work, the authors used the Feynman Path Integral formalism to quantify the effects.
In the formalism, classical refers to paths which extremize the action whereas non-classical refers to the sub-leading paths which do not extremize the action. Representative paths are shown in the inset of Figure 1. When only classical paths are accounted for, $\kappa$ is manifestly zero. Taking into account non-classical paths which actually represent the deviation from the superposition principle makes $\kappa$ a non-zero quantity. It was found that simulations in \cite{PRL} were equally applicable to both the quantum and classical domain. This was followed by an analytic version of the work \cite{Scirep}. In \cite{draedt}, Finite Difference Time Domain (FDTD) simulations of $\kappa$ were carried out which showed that the boundary conditions play a crucial role in the classical electromagnetic domain.\\
The importance of boundary conditions was first pointed out by Yabuki \cite{yabuki} in his theory work involving path integrals and double slits. However, although there has been a lot of theoretical interest in this problem over the last few years, experiments to measure this quantity \cite{usinha,sollner,park,PRA,gregorone,gregortwo,arndt} have been unable to report a non zero value for $\kappa$ due to the error contribution being much larger than the expected non-zeroness of $\kappa$. Earlier experiments on measuring the non-zero Sorkin parameter were focused on quantifying the accuracy of the Born rule for probabilities in quantum mechanics. According to \cite{sorkin}, a non-zero Sorkin parameter would imply violation of the Born rule. However, \cite{draedt, PRL, Scirep} have shown in recent times that the Sorkin parameter was never meant to be zero anyway due to boundary condition considerations, thus making it a less efficient measure for the Born rule, especially in slit based interference experiments. In this paper, we report measurement of a non-zero Sorkin parameter for the first time in the microwave domain using a triple slot experiment which is much above the error bound. We find that boundary condition arguments related to the correct application of the superposition principle are sufficient to explain the non-zero value and thus exemplify that Born rule need not be violated for a non-zero Sorkin parameter. 

\begin{figure*}
\centering
\includegraphics[width=0.9\linewidth]{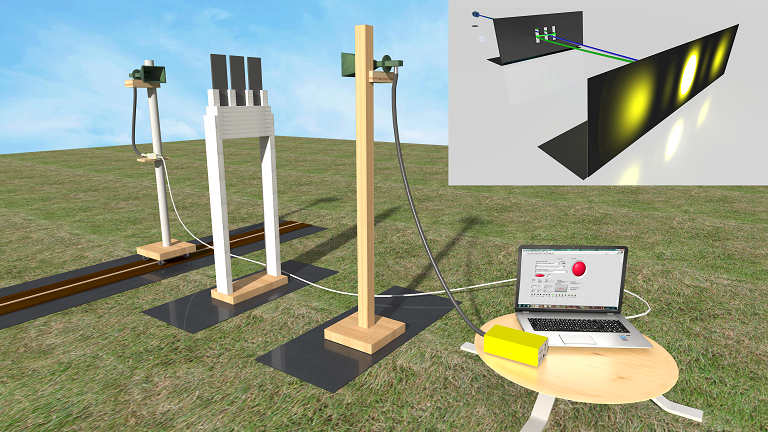}
\caption{A schematic of the experimental set-up. The green antennas on either side are pyramidal horn antennas which act as the source and detector of electromagnetic waves respectively. The detector antenna is placed on a moving rail to enable measuring of diffraction patterns. The three slots are placed between the source and the detector. The inset shows a triple slit schematic where the blue line is a representative classical path and the green line a representative non classical path in path integral formalism.}
\label{setup}
\end{figure*}

Another recent work \cite{boyd} has shown the measurement of a non-zero Sorkin parameter in a completely different experimental scheme using fundamentally different amplification techniques for an enhanced effect. Their triple slit experiment was done using single photons of 810 nm wavelength by enhancing the electromagnetic near-fields in the vicinity of slits through excitation of surface plasmons in the material used for etching the slits. Thus, they enhanced the Sorkin parameter by using near field components of the photon wave function and material induced effects. On the other hand, not only have we have observed non zero Sorkin parameter which is purely due to length scale dependent boundary condition effect on the superposition principle,  our experiment is also in a completely different wavelength domain. As per \cite{Scirep, PRL}, the Sorkin parameter is a length scale dependent parameter and we have used this to our advantage by designing an experiment which uses the microwave length scale to predict and observe a large parameter, much above the error bound. Our experiment is also unique in being a tunable experiment in which we have used obstructions to the slit plane to minimize and finally cancel the effects due to the non classical paths and then remove the obstructions to bring the effect back. Thus, we can increase and decrease the effect due to the non-classical paths at will, making this a definitive proof of their existence than for instance \cite{boyd} which could only see the effect due to all possible paths (and not have any control on the presence/absence of non classical paths) but did not have the ability to tune them at will. Our experiment thus brings forth the reality of Feynman paths in a classical domain.

We have performed a precision triple slot experiment on an open field in the centimetre wave domain using pyramidal horn antennas as sources and detectors of electromagnetic waves and specially manufactured composite materials as microwave absorbers to provide us with the slots. The open field chosen was in a remote observatory which is free from spurious rf noise, man made noise and interference as well as negligible ground reflections, almost mimicking an anechoic chamber. 
The measured graphs of $\kappa$ as a function of detector position have good agreement with theoretically simulated plots using the Method of Moments (MOM is a 3D simulation technique in which exact horn detector and slot material parameters can also be simulated). These results demonstrate the importance of taking proper boundary conditions into consideration while applying the superposition principle in slit/slot based interference experiments.This is essentially a boundary value problem and need not need a quantum mechanical explanation per se. They also describe experimental situations in which the non-zeroness of the Sorkin parameter need not be a good measure for Born rule violation. Our experiment is actually testing the length scale dependent boundary condition effect on the application of the Superposition principle in interference experiments. Thus, it also serves as a guide to future experiments intending to test Born rule violations which need to be properly designed to minimize these effects. They also exemplify a situation in which {\it not} just the  classically dominant paths in the Path Integral formalism has been used to explain experimental observations. They thus bring forth the experimental reality of the winding paths in Path Integrals.

\begin{figure*}
\begin{center}
\includegraphics[width=0.7\linewidth]{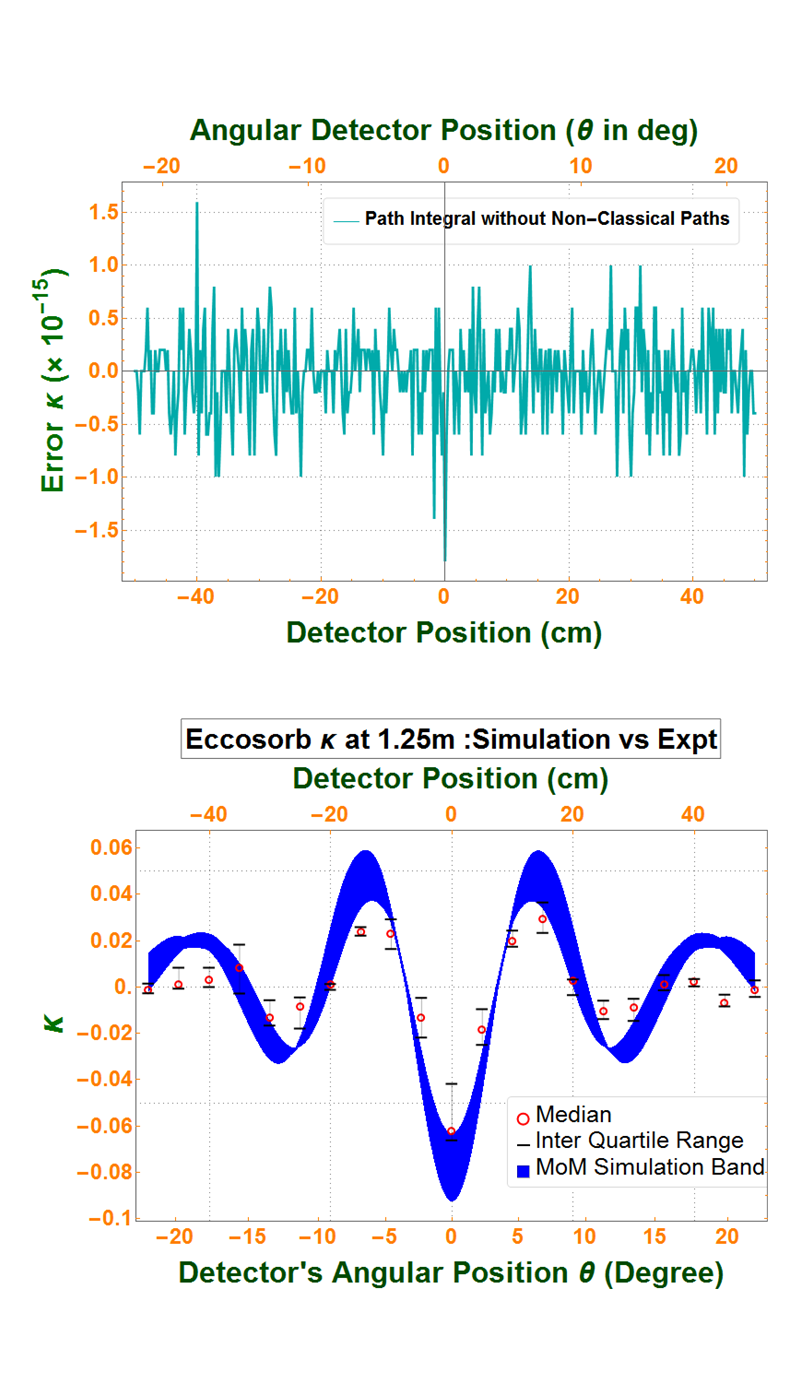}
\caption{On top is a simulation result which is a plot of $\kappa$ as a function of detector's angular position (which is the angle subtended at the detector plane by the centre of the slot plane) when the superposition principle is incorrectly applied. We have obtained this by taking into account only classical paths in the path integral formalism whereby $\kappa$ is manifestly zero. The plot below is a representative experimentally measured $\kappa$ as a function of detector's angular position. The red markers represent the median $\kappa$ at each position. The black lines denote the interquartile range with outliers removed. A few detector positions had one or two outliers. The blue band represents the theory band obtained from MOM based simulation. To create the theory band, we simulated $\kappa$ for the experimental parameters taking into account uncertainties in the parameters. Major contributions to the band came from antenna probe wire height, distance from backplate, interslot distance as well as the material refractive index based uncertainties. }
\end{center}
\end{figure*}

\begin{figure*}
\begin{minipage}{.7\linewidth}
\centering
\includegraphics[width=0.7\textwidth]{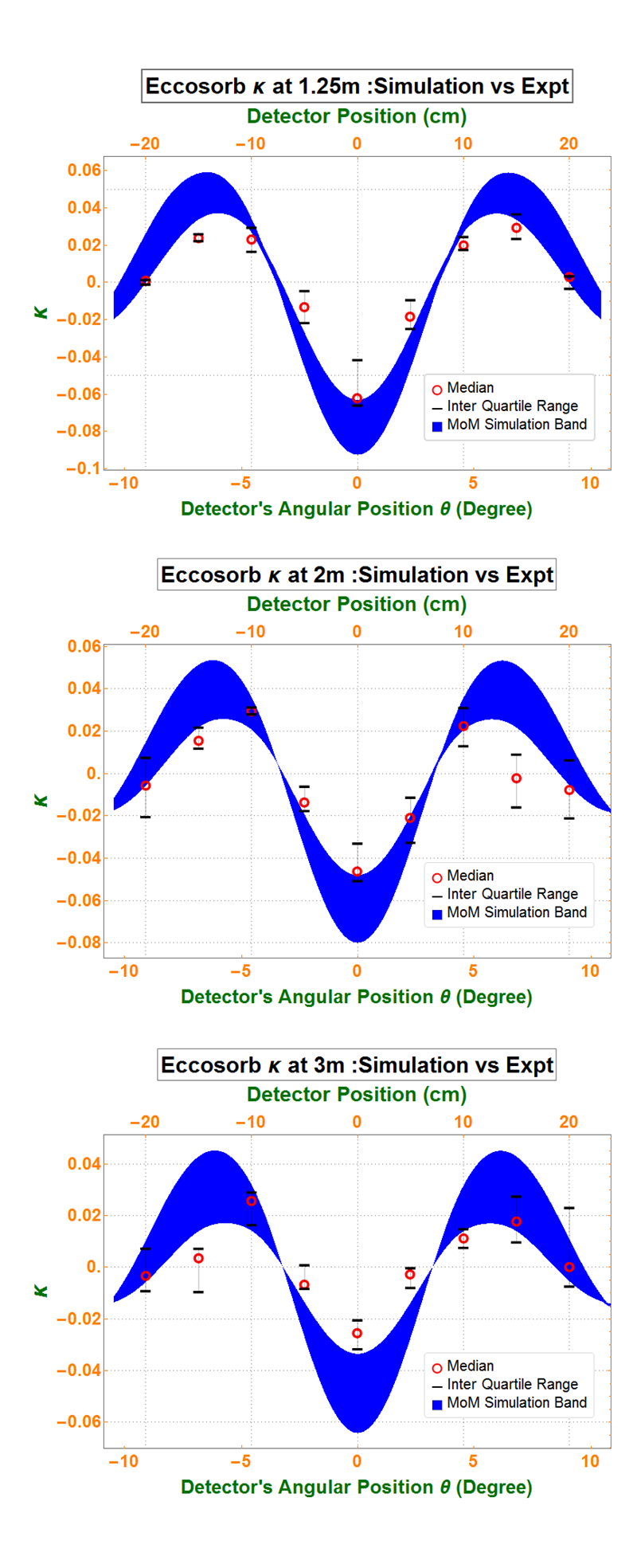}
\caption{The plot on top is for source-slot plane distance of 1.25 m. The middle plot is for source-slot plane distance of 2 m. The bottom plot is for source-slot plane distance of 3 m. Slot plane - detector plane distance is kept 1.25 m in all three cases. As can be seen, experiment and theory match very well in all cases. Theory predicts a drop in the $\kappa$ values with increasing source-slot plane distance which is corroborated by experiment. As distance increases, the noise remains similar but signal drops making signal to noise go down which results in generally bigger error bars at larger distance. }
\end{minipage}
\end{figure*}

\section{Results}

In this paper \cite{talkone, talktwo}, we report results of a triple slot experiment. As the dimensions are macroscopic (centimetre scale), for a commensurate slit based experiment,  in order to consider a suitable outer box for simulations, we would need to etch the slits in an absorbing layer which needs to be several metres long. This is both practically and economically prohibitive. On the other hand, having absorbing slots surrounded by free space it is much simpler to mimic infinitely large boundaries. 

The definition of $\kappa$ in case of the triple slot experiment becomes:
\begin{equation}\label{kappaold}
\kappa = \frac{ p_{BG} - (p_{ABC}-p_{AB}-p_{BC}-p_{CA}+p_A+p_B+p_C)}{Max(p_{BG})}, \
\end{equation}
where $p_{BG}$ is the magnitude of the Poynting vector at a certain detector position due to the horn source (in experiment, it is the measured power value) and $Max(p_{BG})$ is the maximum value of the same. $p_{x}$ stands for magnitude of the Poynting vector at a certain detector position due to the presence of slot combination $x$ where $x = A, B, C, AB, BC, AC, ABC$. Further details on the above definition are included in \cite{supple}.

\subsection{Experimental details}

Figure \ref{setup} shows the details of the experimental set-up. We have a pyramidal horn antenna acting as a source of electromagnetic waves at 5 cm wavelength. These waves are incident on slots which are 10 cm wide having an inter-slot distance of 13 cm (centre to centre). The slots are composite structures which consist of two layers of specially manufactured materials which act as near perfect absorbers of microwave frequency (Eccosorb SF6.0) sandwiched with an aluminium layer in between to enable even more perfect absorption especially of back reflected beams which may make their way back to the source antenna from the detector. We have done some rigorous analysis of such back reflection and concluded that they do not affect our experiments \cite{supple}. The detector antenna is also a horn antenna similar to the source but housed on a moving rail to enable collection of data as a function of detector position. We use a high frequency power probe to record power values for the different combinations of slots required for measuring $\kappa$ as a function of detector position. The measurement involves eight separate experiments which measure the individual contributions to equation \ref{kappaold}. A ninth measurement involves measuring with source off and detector on to confirm that the antenna does not pick up any comparable signal from unknown emitters in the environment. This was several orders of magnitude lower than the measured values with the source on leading to a very low stray signal level. 
In the Methods section, we have included technical details on how we aimed at achieving perfect alignment which plays a crucial role in a precision experiment like this one.

\subsection{Experimental results}

Figure 2 shows a representative plot of $\kappa$ as a function of detector position at a source detector separation of 2.5 metres (1.25 metres between source and slots plane and same distance between slot plane and detector plane). The detector movement is controlled using a 10 RPM DC motor. At each detector position, all eight contributions to equation \ref{kappaold} are measured by placing and removing slots as required before the detector is moved to the next position. This way, we ensure that the $\kappa$ value at a certain detector position is measured on a short time scale and fluctuations due to the environment do not affect the individual contributions in such a way that it can affect the value of $\kappa$ measured. We have ensured that our measurement time scale is sufficient by actually measuring an antenna radiation pattern as a background measurement before measuring the pattern due to a particular slot combination. These background patterns overlap during the $\kappa$ measurement time indicating that our noise fluctuation time scale is much longer than the measurement time. The background corresponding to a certain slot combination is taken to be the average of the background value measured before the combination and after. The formula for $\kappa$ measured experimentally is modified as follows.

\begin{eqnarray}
\kappa =  \gamma (P_{ABC}-P_{AB}-P_{BC}-P_{CA}+P_A+P_B+P_C) \ 
\label{eqn:new}
\end{eqnarray}
 
 where $P_\alpha = \frac{p_{BG\alpha} - p_\alpha}{p_{BG\alpha}}$, $\alpha = A, B, C,...ABC$ slots being present and $p_{BG\alpha}$ refers to the background corresponding to each combination. $\gamma = \frac{p_{BG(at~ x = x_D)}}{Max(p_{BG}}$, $x_D$ being a certain detector position. As can be seen, at the position corresponding to maximum of the background, which is usually the centre of the radiation pattern, $\gamma = 1$ and the equations \ref{kappaold} and \ref{eqn:new} become equivalent. By defining individual background contributions to the different terms in equation \ref{kappaold}, we can take care of varying source power, if any, between combinations. The background value is measured using a reference detector and also by averaging between background values measured before and after each combination which turn out to be equivalent in our case.\\
 For each combination, 3000 data points are collected, the median of which contributes the measured power value. Thus each $\kappa$ value has 3000 measured values for each combination. The number 3000 was arrived at after sampling for both lower and higher number of data points (it was found that the $\kappa$ values converge to the same value for 2000 data points per combination itself so we decided to measure 3000 data points for each combination to overcompensate). 10 measurements of $\kappa$ were done at each detector position and the median value has been chosen as the representative value. The errors for each value have been represented by box plots \cite{box}. Further details on choice of measurement statistics as well as error analysis are given in the Methods section. We have also randomised the order in which slot combinations are measured and ensured that $\kappa$ remains constant.

Figure 3 shows measured $\kappa$ as a function of detector position for three different source-detector distances. This was done to ensure that the match between theory and experiment persists even on changing some changeable parameters in the experiment.

\subsection{Killing $\kappa$ with a baffle}

In order to have an independent check on the fact that $\kappa$ as measured is not due to some unaccounted for errors, we have added absorbers perpendicular to the slots (in between the slots) to in principle ``slowly kill" the effects due to non classical paths (see Figure 4). \cite{PRL} had postulated that the paths which cross the slit plane twice i.e. the hugging paths would have maximum contribution to $\kappa$. By placing such perpendicular baffles and measuring $\kappa$ as a function of increasing baffle size, we find that  $\kappa$ decreases with increase in baffle size and show that it is indeed the hugging paths that contribute maximally to non-zero $\kappa$. Figure 5 shows $|\kappa|$ at central detector position as a function of increasing baffle size. This is definitive alternative proof that what we are observing is a real physical effect and not a result of some unaccounted for error. This makes ours a tunable experiment where the baffle size gives the tunability parameter. In the absence of baffle, both classical and non-classical paths contribute to $\kappa$. As the baffle size increases, the contribution due to non-classical paths diminishes and finally becomes zero (which is observed by $\kappa$ becoming zero). Thus, the baffle experiment demonstrates the true effect due to the exotic non-classical paths and brings forth the reality of Feynman paths in a classical domain. 

\begin{figure*}
\centering
\includegraphics[width=0.75\linewidth]{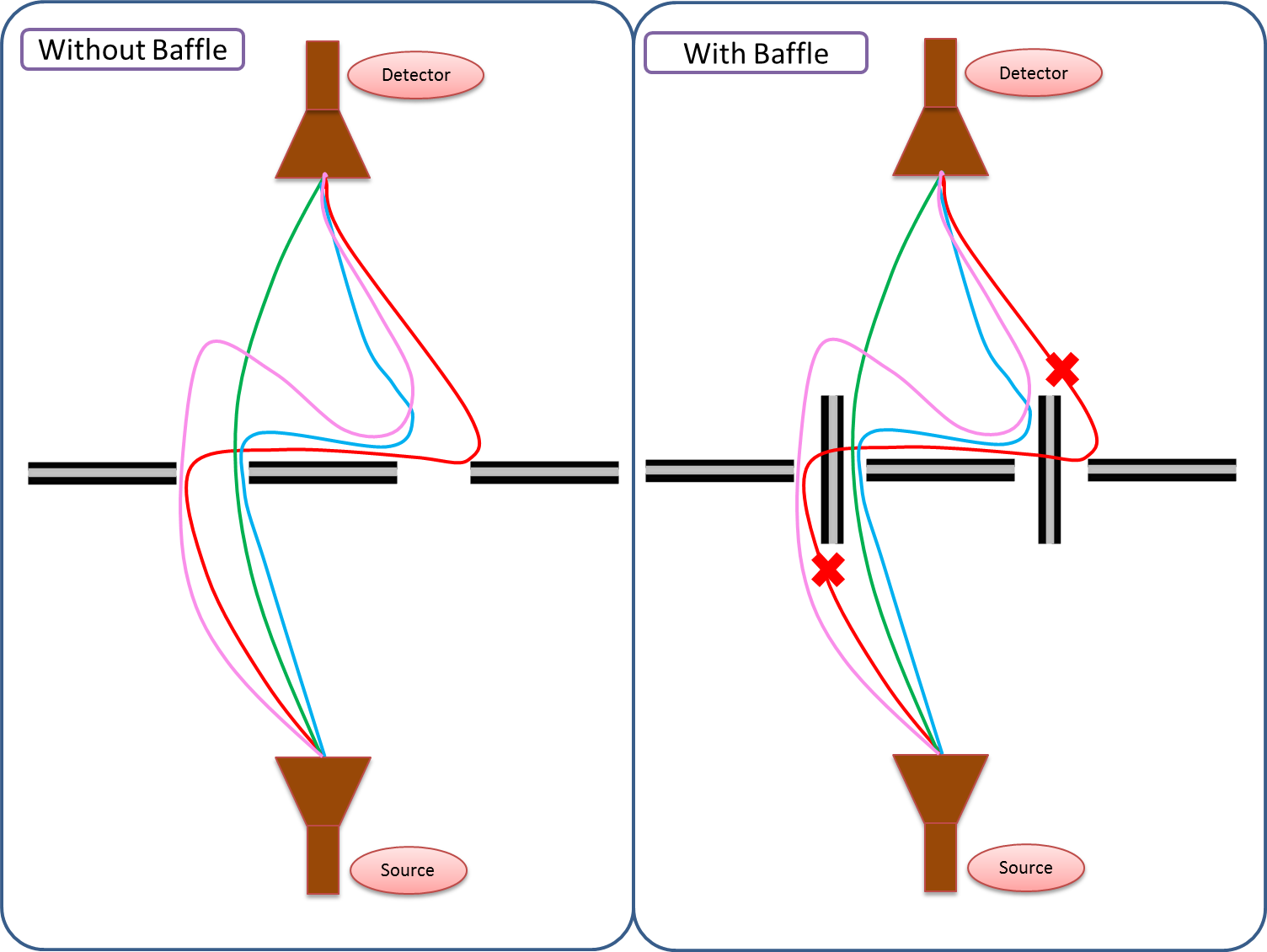}
\caption{Red lines indicate class of non classical paths that get blocked by baffle, Blue lines indicate those that are still allowed and may contribute to $\kappa$, Pink lines indicate those that get suppressed with increase in baffle size, Green lines indicate classical paths.}
\end{figure*}

\begin{figure*}
\centering
\includegraphics[width=0.7\linewidth]{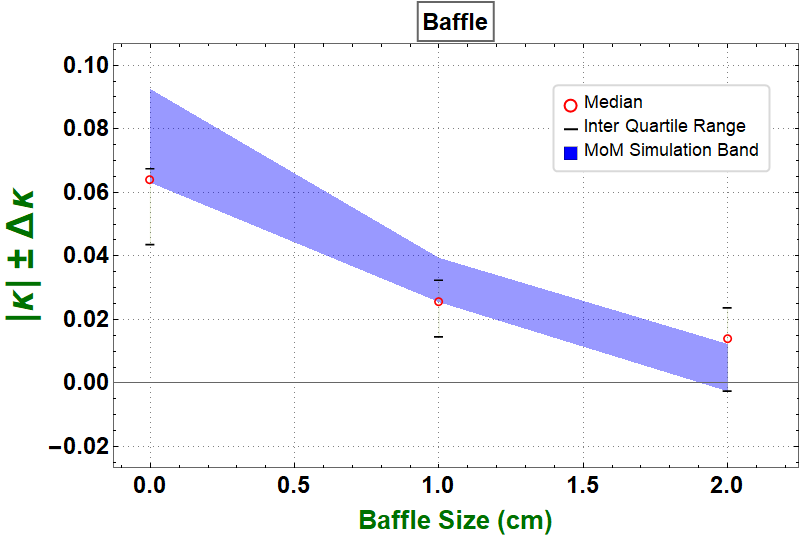}
\caption{$|\kappa|$ at central detector position as a function of baffle size. Red markers represent medians of 10 measured $|\kappa|$ values for different baffle sizes, black lines denote the interquartile range, shaded blue region is the theory band from MOM simulation. We have used  $p_{BG}$  at the central detector position as the normalization factor. In order to plot $|\kappa|$, we take the absolute value of the median but keep the relative distance of interquartile range.}
\end{figure*}

\begin{figure*}
\centering
\includegraphics[width=0.7\linewidth]{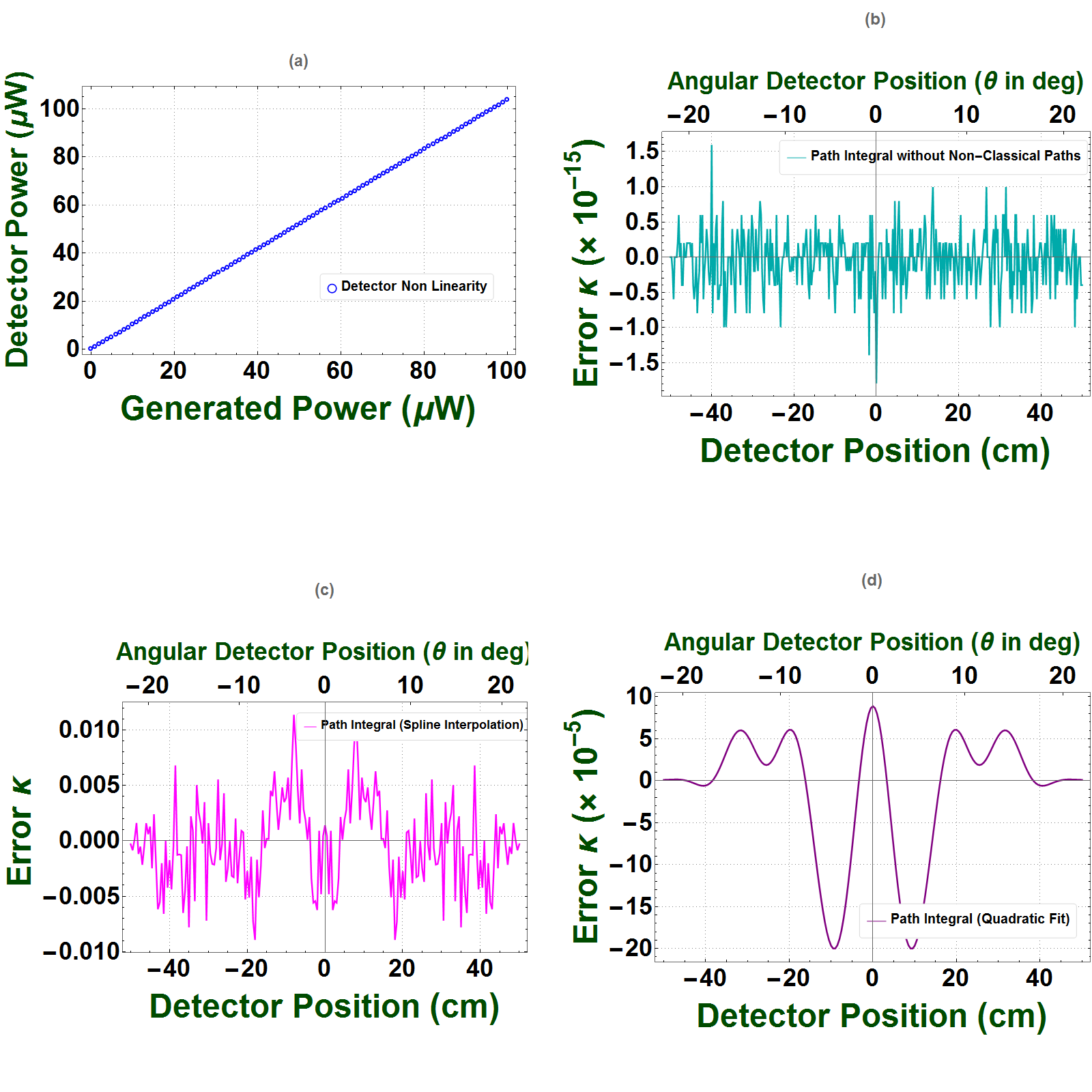}
\caption{a) measured power values versus input power b) $\kappa$ generated from classical paths only c) $\kappa$ generated from spline interpolation method d) $\kappa$ generated from polynomial fit.}
\end{figure*}

\begin{figure*}
\centering
\includegraphics[width=0.7\linewidth]{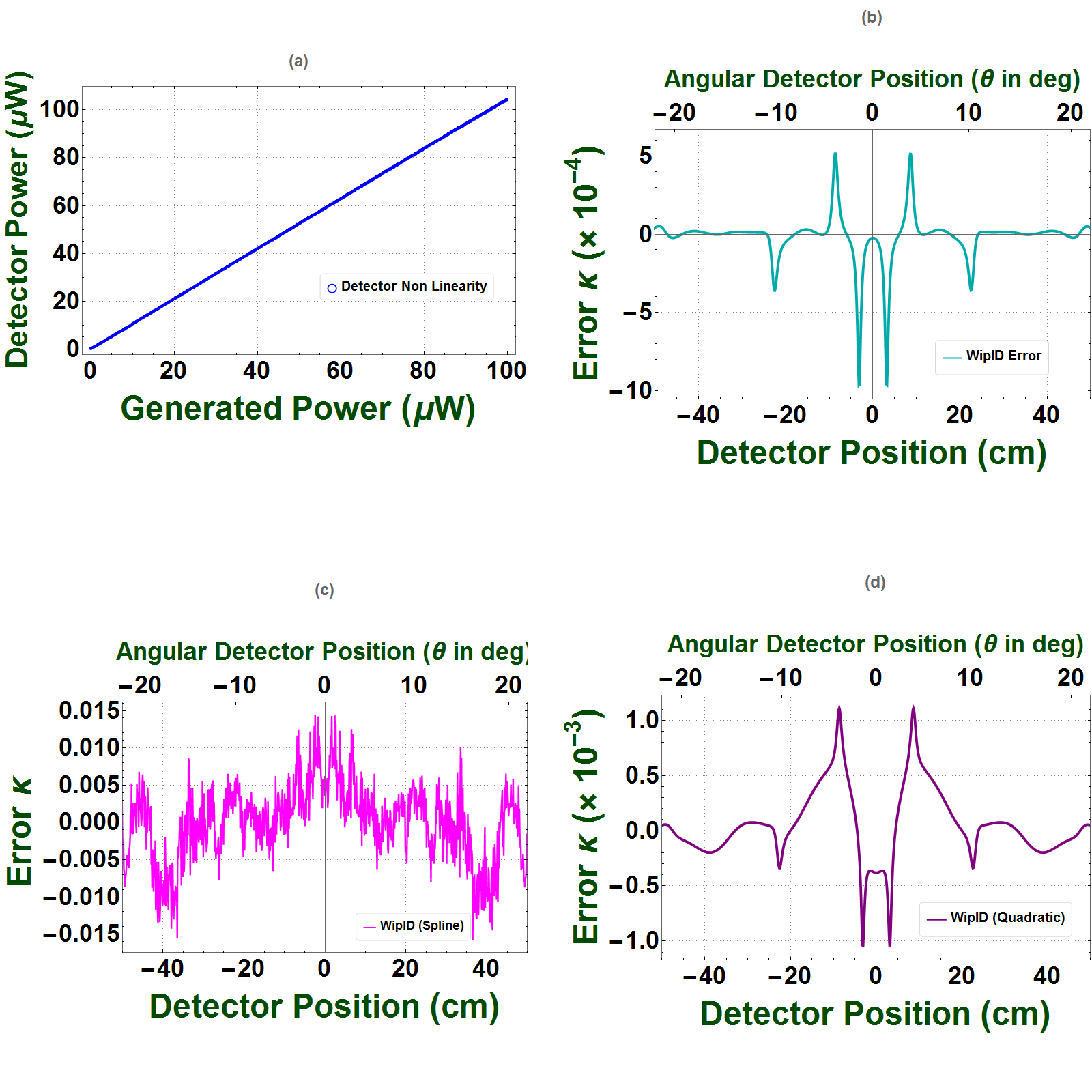}
\caption{a) measured power values versus input power b) $\kappa$ generated from the naive application of the superposition principle using MOM c) $\kappa$ generated from spline interpolation method d) $\kappa$ generated from polynomial fit.}
\end{figure*}

\section{Discussion}

In addition to the above experiments, we have simulated the effect on $\kappa$ from changing detector size and also done simulations using path integral formalism and FDTD and compared the two as shown in the Methods section.

\subsection{Detector non linearity analysis}

One of the main errors which can lead to observation of a non-zero $\kappa$ will be the non-linearity if any of the detector. If a detector behaves non-linearly with increase in incident power, then the quantity in equation \ref{kappaold} will automatically be a non-zero just from such errors. We have done detailed analysis of detector non-linearity and found that our measured $\kappa$ cannot be explained by any such non-linearity, effects due to which happens to be much below the measured $\kappa$. In our analysis, we have derived a non-linear function for the detector using both spline interpolation method as well as polynomial fit and derived the resultant $\kappa$ from this function. We have found the $\kappa$ value so derived to be much lower than the measured $\kappa$ thus indicating that the non linearity effects do not play a major role in our experiment. 

Figure 6 shows four plots.

Plot (6a) shows measured power values in our Agilent power probe with an Agilent signal generator acting as a source. If the detector is perfectly linear, then the measured value will be exactly the same as the input value. However, no detector can be linear upto arbitrary accuracy. We have used this plot to generate non linear functions from both spline interpolation method as well as polynomial fit.
Plot (6b) shows the $\kappa$ that is generated using only classical paths in the path integral formalism. As is expected, $\kappa$ is identically zero ($~10^{-16}$ which is the accuracy of our solver, i.e. float precision for Mathematica.). 
We have used the power values that lead to this zero value for $\kappa$ and feeding them to the non linear function generated above, derived the power values that would have been measured. The measured value will vary from the input value due to various non linearity effects.
Plot (6c) and Plot (6d) show the resultant $\kappa$ as a function of detector position. These values represent what we define to be $error~ \kappa$.  What we are verifying in this experiment is the deviation from the naive application of the superposition principle. While the naive application gives us an expected zero value for $\kappa$, the correct application brings forth the non-zeroness. Following references [6] and [8], we define $\kappa$ only in the presence of ``classical" paths or in other words when the Superposition principle is naively applied. This is expected to be zero in ideal theory. However, in case of real experimental/simulation scenarios which involve non-ideal conditions, this quantity can be a non-zero. One has to appreciate that this non-zero is simply due to different sources of error as the case may be and has nothing to do with the correction to the application of the Superposition principle which is a``real" non-zero as opposed to an error bound. We call such an error bound``error $\kappa$." This quantity derived simply from the terms involving classical paths comes in very handy as it tells us whether some source of error has  a competing effect with the actual non-zero value. Thus if error $\kappa$ is lower order in magnitude than actual $\kappa$, we need not worry about a particular error playing a dominant role in explaining the non-zeroness of $\kappa$. In case of Plot (6c) and (6d), as they have been generated from taking into account only the contribution from the classical paths in path integral formalism, they should have been zero. However, non linearity effects make them non-zero.

Figure 7 shows a similar analysis done using WIPL-D (our MOM solver) in which the accuracy of the solver is $~10^{-4}$.

Two very important points should be noted here. \\
\begin{itemize}
\item The non linearity effects captured here reflect the maximum non linearity that can affect our experiment which is of course not representative. Even in this worst case scenario based simulation, the values of $\kappa$ are many times smaller for the interpolation method and two to three orders of magnitude lower for the polynomial fit. Thus, they do not in any way explain the results obtained in the experiment.
\item Plot (a) captures the non linearity not only due to the power probe but also the source signal generator itself. There is no trusted device that one can assume is perfectly linear and use as a source such that only the measurement device non linearity can be captured. In our experiment, the source is used at a constant power and thus non linearity due to the source does not affect us. The effective non linearity seen in our experiment is thus lower than what we have been able to estimate. The issue of a trusted device also existed in previous work \cite{gregorone} as there the attenuator was assumed to be a trusted device.
\end {itemize}

\subsection{Approximations sometimes made in astronomy}

The experimental result reported in this manuscript has several implications in optics as well as related areas of research like Radio Astronomy. In the latter, the community is divided in the sense that while some work in available literature seems to take into account boundary condition effects \cite{radioone, radiotwo}, there are others which seem to ignore them \cite{radiothree, radiofour}. We have explored this application in further detail and found that by taking relevant parameters from such experiments \cite{radiofour}, we get $\kappa$ to be of the order of $10^{-2}$ which is definitely not ignorable any more considering that here we are reporting an experiment where we have successfully measured the quantity much above the error bound. There are some applications in this field where the naive application of the superposition principle is routinely used, for instance in calculation of array factor \cite{radiofive} as well as in estimation of effects of badly behaving antennas in an array configuration. We find that for very large arrays, such approximations may hold upto a point; however, the gain calculated using correct boundary conditions (MOM simulations) gives much better match than array factor at higher angles. The validity of the approximations is inversely proportional to the array size and is also dependent on whether the absolute power value is of concern (in which case boundary conditions play a big role as opposed to normalization with bright sources in the sky). In any event, our current experimental results and calculations using radio astronomy parameters tells us that these boundary conditions will play a crucial role in future experiments on precision astronomy where errors from other sources would have been suitably minimized. Further details can be found in the Methods section. 

\subsection{Conclusion}

Being one of the first non-zero detections of the normalized Sorkin parameter, our experiment vindicates the recent claims of different theory papers that the superposition principle when applied to slit based interference experiments needs a correction term originating from the difference in boundary conditions presented between multiple slits opened all at once compared with a summation of the effects from the slits being opened one at a time \cite{PRL, Scirep, draedt, yabuki}. This correction term has been expressed in terms of the contribution from non-classical paths in the Feynman Path Integral formalism \cite{PRL,Scirep}. Ours being a tunable experiment, we are able to increase and decrease the effect due to these non-classical paths at will which no other experiment has done so far, thus bringing forth the reality of Feynman paths in a classical experiment without any ambiguity.\\
The non-zero value of the correction term obtained in this experiment is well explained by the correct application of the superposition principle and does not need Born rule to be violated and thus has immediate implications for future experiments aimed at testing the Born rule. Our experiment is done in a microwave length scale domain and uses geometry effects to observe non-zero Sorkin parameter, thus providing a benchmark for future experiments. Any such experiment has to be carefully designed to minimize the length scale dependent effects on the application of the superposition principle.This is a fundamentally important experimental result and is expected to play a major role in the quest for genuine post quantum higher order interference. Higher order interference was initially discussed by Sorkin \cite{sorkin} in the framework of Quantum Measure Theory. Recently, a lot of theoretical thrust has gone towards developing what are called generalized probabilistic theories \cite{barnum} which actually require as a postulate that higher order interference be zero. Our work which puts a non-zero bound on higher order interference from pure length scale dependent boundary condition arguments then naturally raises the following questions: Does this non-zeroness affect such post quantum theories \cite{arxivselby, walther} and if so, how? What will be the far-reaching implications now that the Sorkin parameter has been proven to be non-zero within the realms of quantum physics and classical electromagnetism? Finally, a question of much practical significance arises. Since genuine higher order interference is being investigated as a possible resource in computation \cite{NJPselby, phaseselby}, how will the experimental verification that higher order interference turns out to be non-zero affect such research directions? \\

\section{Methods}

\section*{Precision alignment}

One of the most crucial steps which enables us to measure a convincing non-zero for $\kappa$ involves precise alignment of the various components in the experiment. While there is alignment at a basic level, there is also finer alignment using dedicated tools. 
\begin {itemize}
\item The first condition to be ensured is perfect levelling of the ground. A spirit level is used at various points on the ground to check the ground level and all unevenness is filled with sand. Once preliminary levelling is achieved, we place a marble-like stone on the ground to ensure further smoothness. These stones need to be settled into the ground using water so that once set, they do not sag any further. 

\item The experimental set-up consists of a rail, a motor for horn antenna detector movement, two horn antennas, slot stand and slots. The source horn antenna is connected to an Anapico signal generator which generates microwaves at 6 GHz. The receiver horn antenna is connected to a high frequency power probe from Agilent. The data acquisition is done by using Labview.
\item The next alignment involves alignment of transmitter and receiver horn antennas.
\item The transmitter horn antenna is fixed while the receiver antenna moves on a rail. For initial alignment, the receiver antenna is placed at the centre of the rail directly in front of the source antenna. It is ensured that the two antenna centres coincide with each other. This is done using a plumb line.
\item Once the perpendicular alignment is done, one needs to ensure that the distance between the edge of the rail on either side is the same from the source. This is done using a Laser Distance metre (LDM).
\item Next, the slot stand (made of high density thermocol which is separately checked to be a perfect transmitter for the 6 GHz microwaves upto the desired accuracy)is placed between the source and receiver. The distance between the source antenna probe wire and the receiver antenna probe wire is 2.50 metres and the slot stand is exactly 1.25 metres from each. The distances are again ensured using LDM.  
\item Water level is used to ensure that the height of both the source and receiver antennas are 1.75 metres from the ground.

\item Figure 8 shows the final set up which is housed in an appropriate tent for protection against wind and rain.

\begin{figure}
\centering
\includegraphics[width=0.5\linewidth]{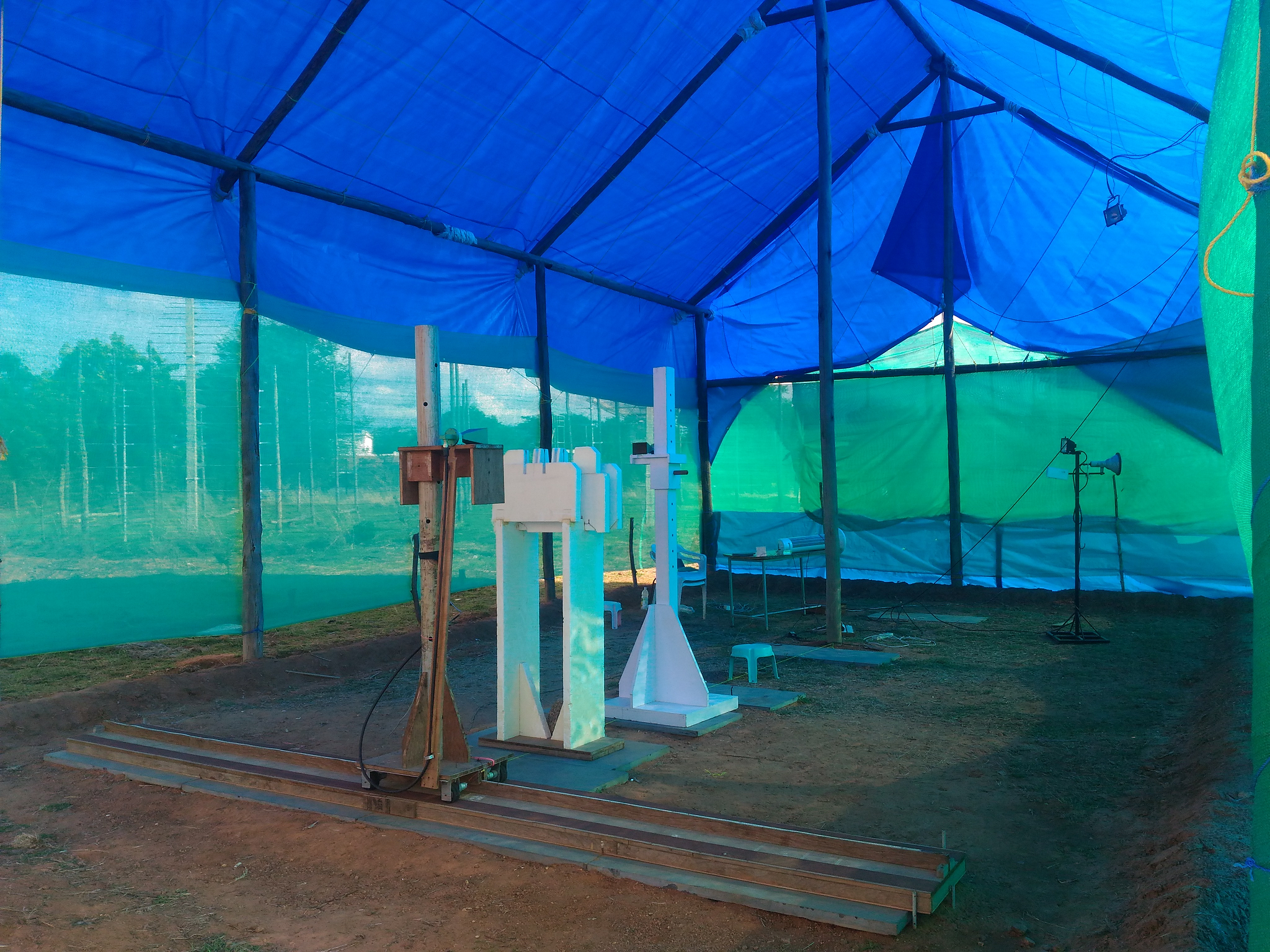}
\caption{Final experimental set up which is housed in a tent}
\end{figure}

\item One of the considerations that plays a role in experiments which are performed in open field conditions is the possibility of reflection from the ground affecting the source radiation. Other than the back reflections from metal and other structures on the field (which we have discussed in detail in the supplementary material), we also need to confirm that reflections from the earth's surface do not cause any difference in measured power. One can reduce the effective reflection from the ground by raising antenna height to an extent that these reflections do not play any role. Figure 9 shows a plot of measured power vs height of source and receiver antenna at different source powers (0 dBm, 10 dBm, 15 dBm) when the source and receiver antenna are at a distance of 8 metres from each other. As one can see, beyond 145 cm height of the antennas, there is no appreciable change in power measured. This implies that beyond this height, there is no relevant change caused due to specular reflection. Our experiment was conducted at source and receiver antenna heights of 175 cm and at a smallest source-receiver distance of 2.5 metres where the specular reflection component will be even more negligible.

\begin{figure}[hb]
\centering
\includegraphics[width=0.9\linewidth]{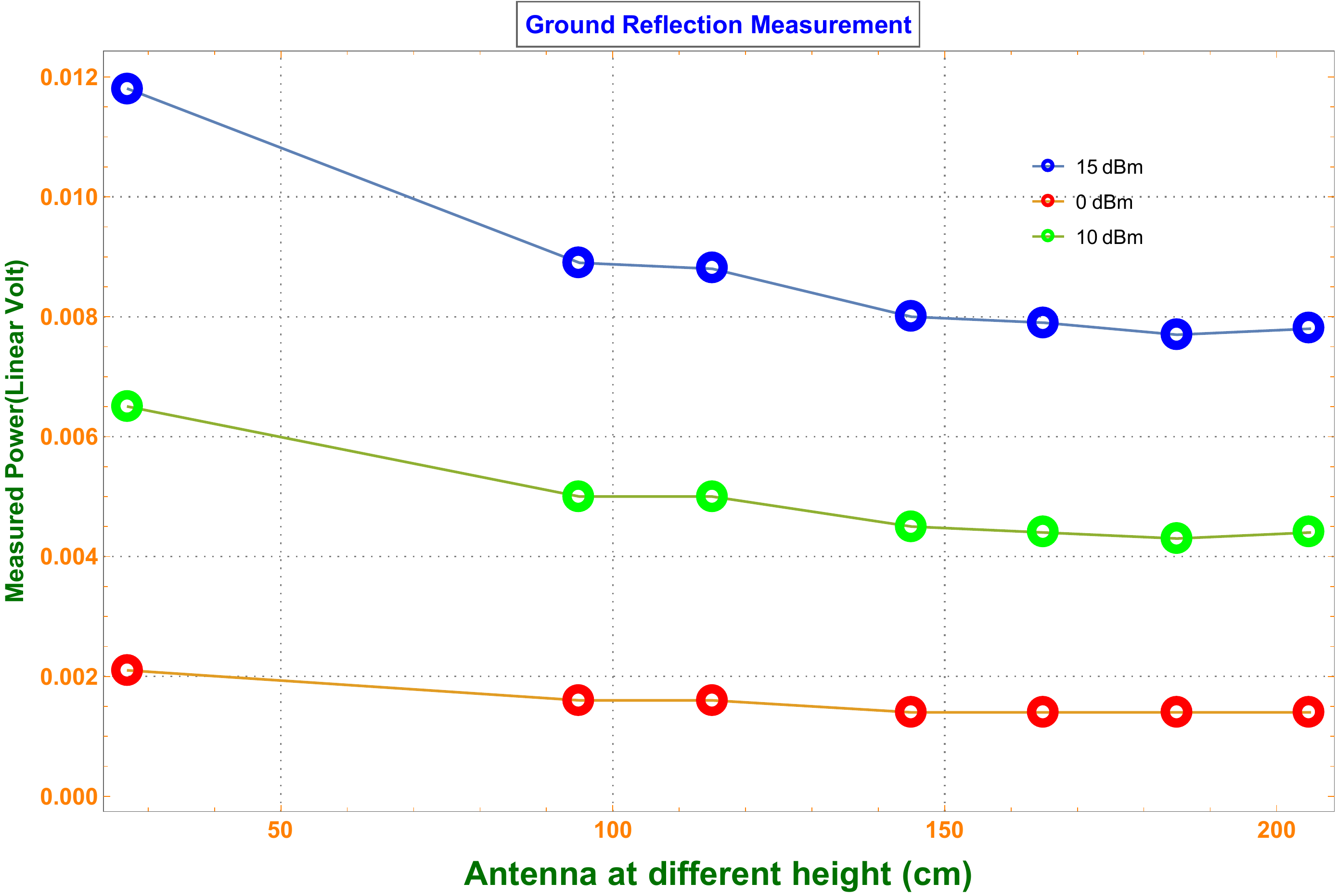}
\caption{Measured power values as a function of antenna height for source-receiver distance of 8 metres. The values saturate to almost the same value beyond 145 cm.}
\end{figure}
\end {itemize}

 In order to substantiate this point further, we include below a geometry based argument which proves that ground reflections do not play a role in our experiment.
Consider the two horn antennas 2.5 m apart at a height of 1.75 m from ground as shown in Figure 10. From laws of reflection, both source and detector antennas will see the waves reflected from the ground at a distance of 1.25 m from it. Since both the antennas are at 1.75 m above the ground level, the angle subtended by the ray that gets detected after ground reflection from the line joining the two detectors is
\begin{equation}
\arctan\left(\frac{1.75 \ m }{1.25 \ m}\right) = 54.46 \degree
\end{equation}

If we consider that the direct beam has a unit gain from the antenna and factor in the gain at $54.46 \degree$ appropriately, under realistic ground reflection percentages (typically chosen to be $10 \%$), the change in power due to ground reflection (compared to the direct beam) will be about $0.0001$. This is less than experimental errors due to source fluctuations (typically $0.003$) and hence is not a major source of error in the experiment.

\begin{figure*}
\centering
\includegraphics[width=0.99\linewidth]{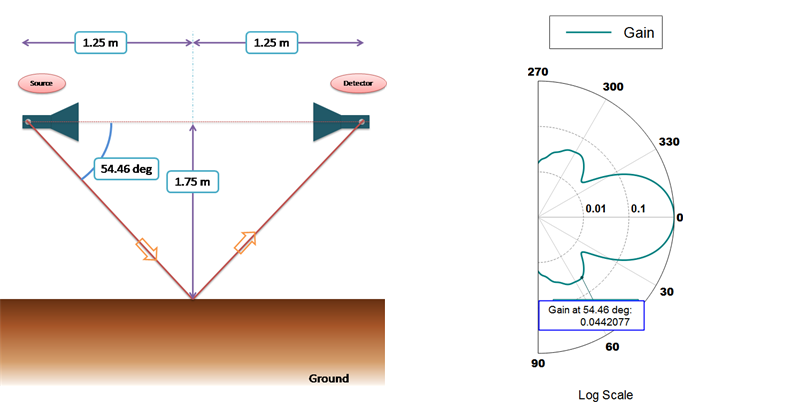}
\caption{Figure on left shows two horn antennas facing one another at a distance of 2.5 metres and at a height of 1.75 metres from the ground. The source and detector antennas both will see waves reflected from the ground at an angle of $54.46 \degree$. Figure on right shows the antenna gain pattern.}
\label{fig:gaintoplin}
\end{figure*}

\section*{Error Analysis}
A precision experiment on an open field may seem like incompatible conditions. Thus, a lot of time and work has gone into ensuring proper representation and accountability of all possible errors. Both random and systematic errors can affect the experiment. A fluctuating source power will add uncertainty to measured $\kappa$ but the mean $\kappa$ will not be affected. On the other hand, a systematic drift in the source power will make different slot combinations experience different effective source powers which in turn will cause a shift in the mean $\kappa$ itself. We have done several hours of source stability analysis and ensured that the error due to fluctuations (typically of the order of $10^-3$) of the source is much below the required precision level. Our source also does not suffer from drift within the time required to measure a $\kappa$ value. This has governed the choice of measurement time for each $\kappa$ value which is typically fifteen minutes. Further precaution has been taken by measuring the background contribution before and after each slot combination and using the average of the two as the representative background value. There are other errors which are unrelated to the source like possible tilt of slot stand, improper fitting on slot in stand etc. We have ensured as near perfect an alignment as possible and repeated the measurement several times to randomize the error further. 
Other than experimental errors, in order to have fair comparison between experiment and theory, we also need to ensure that the theory is not for ideal conditions but in fact takes into account the non idealness and associated uncertainties in different components like length parameters and material parameters. This leads to the generation of the theory band.


We calculate $P_{BG_\alpha}$ by measuring background before and after the slot combination $\alpha$.  Each measurement of background consists of taking 3000 readings (about 45 seconds ). To compute the mean $\mu(P_{BG_\alpha})$ we average over the 6000 readings (combining the readings before and after). To compute the  $\sigma(P_{BG_\alpha})$ we take the standard deviation of 6000 readings. We do the same thing for $p_\alpha$, the slot combination.
From here, we use the formula for $\kappa$ to compute mean $\mu(P_\alpha)$ where all the power values are the mean quantities.

We compute the uncertainty in $P_\alpha$ as follows:
\begin{equation}
\delta P_\alpha = \sqrt{\left(\frac{\partial ( \frac{P_{{BG}_\alpha}- p_\alpha}{P_{{BG}_\alpha}})}{\partial P_{{BG}_\alpha}} \delta P_{BG_\alpha}\right)^2 + \left(\frac{\partial ( \frac{P_{{BG}_\alpha}- p_\alpha}{P_{{BG}_\alpha}})}{\partial P_{\alpha}} \delta P_{\alpha}\right)^2}
\label{sigma}
\end{equation}
In the above equation, we choose to ignore the higher order terms and further assume that background and slot combination powers are independent (in terms of its variation). In the above equation, the partial derivatives are calculated by first taking the partial derivative and then substituting the means. For the variations, the std deviation is used.

This gives us error in each effective slot combination. Then the mean kappa for a data set is computed using commensurate equation [2] in Main Text where in RHS all the quantities are mean for each effective slot combination.  

The error in $\kappa$ is simply computed as
\begin{equation}
\sigma_{\kappa} = \sqrt{\sum (\sigma_{P_\alpha})^2}
\end{equation}

 We measure 10 $\kappa$ values at each detector position.The logic for this choice is as follows. If the standard deviation of the mean of a certain number of readings is lower than the average error bars of each data set, then we can say the experiment is reproducible. In other words, if we take random samplings of a certain number of data sets and calculate the standard deviation for the same, we will find that as the number of data in the random sampling increases, the standard deviation drops and becomes comparable to the individual error bar. In our case, the error bar corresponding to one $\kappa$ value is of the order of $10^{-3}$ and the standard deviation of the random samplings of $\kappa$ values drops to this order after 5 to 6 $\kappa$ measurements at most detector positions. To overcompensate, we have decided to measure 10 $\kappa$ data sets at each detector position. Figure 11 shows a comparison between measuring 10 and 20 $\kappa$ data sets which demonstrates further that 10 is a statistically significant number of data sets to be measured at each detector position.

\begin{figure*}
\centering
\includegraphics[width=0.5\linewidth]{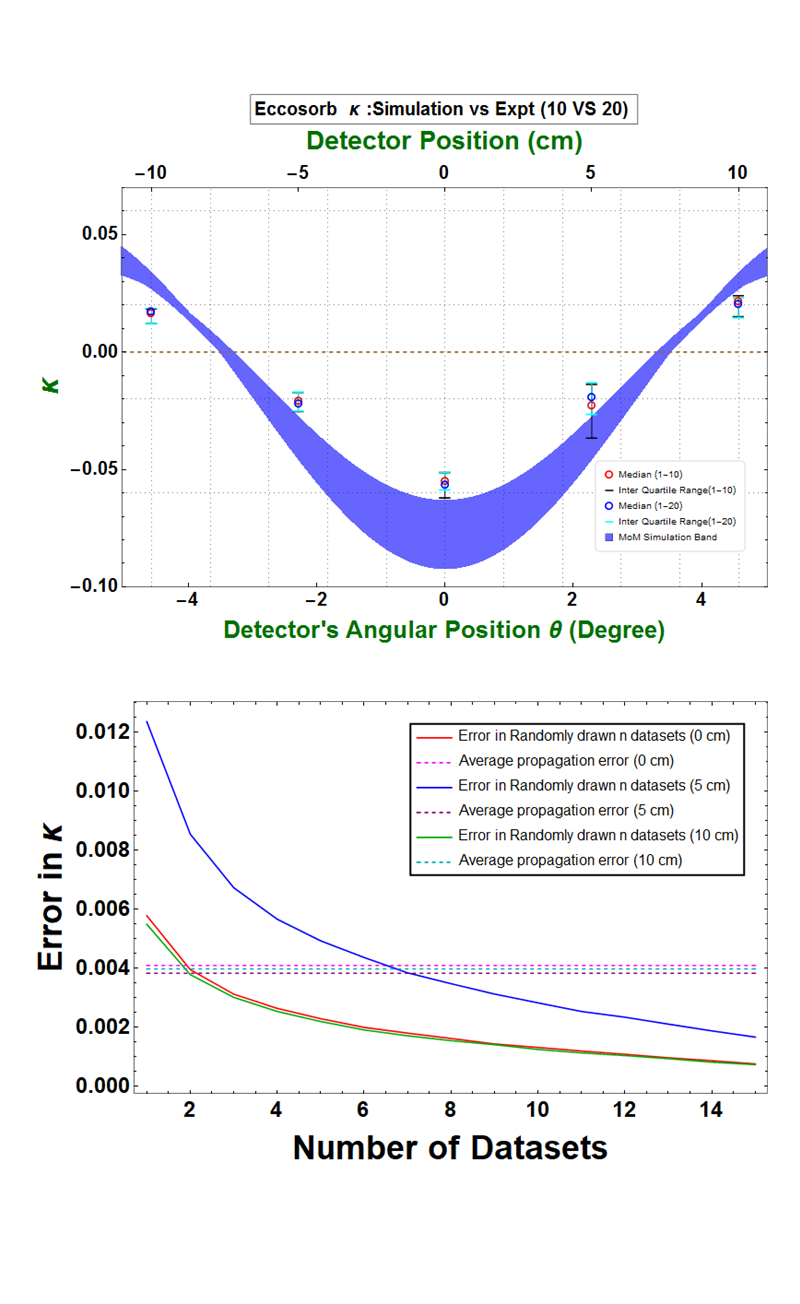}
\caption{Figure on top shows the comparison between measuring 10 $\kappa$ values at 5 detector positions vs measuring 20 $\kappa$ values. As can be seen, measuring more than 10 values does not bring about any appreciable change in median or error bar. The figure below shows how the standard deviation when one takes  random samplings without repetition of a certain number of data sets (here 20) drops as one increases the number of data sets participating in the draw. For instance for the $\kappa$ values measured at the central position (0 cm), even 2 $\kappa$ values make the standard deviation drop below the random propagation error in individual $\kappa$ values which is shown by dotted lines. For the position immediately to the right of the centre (5 cm), the number required for this to happen is higher at 7. However, in none of the positions shown (and otherwise), one needs more than 10 $\kappa$ values for this condition to be satisfied. This implies that 10 is a statistically significant number of data sets to be measured at each detector position. }
\end{figure*}

\begin{figure*}
\centering
\includegraphics[width=0.7\linewidth]{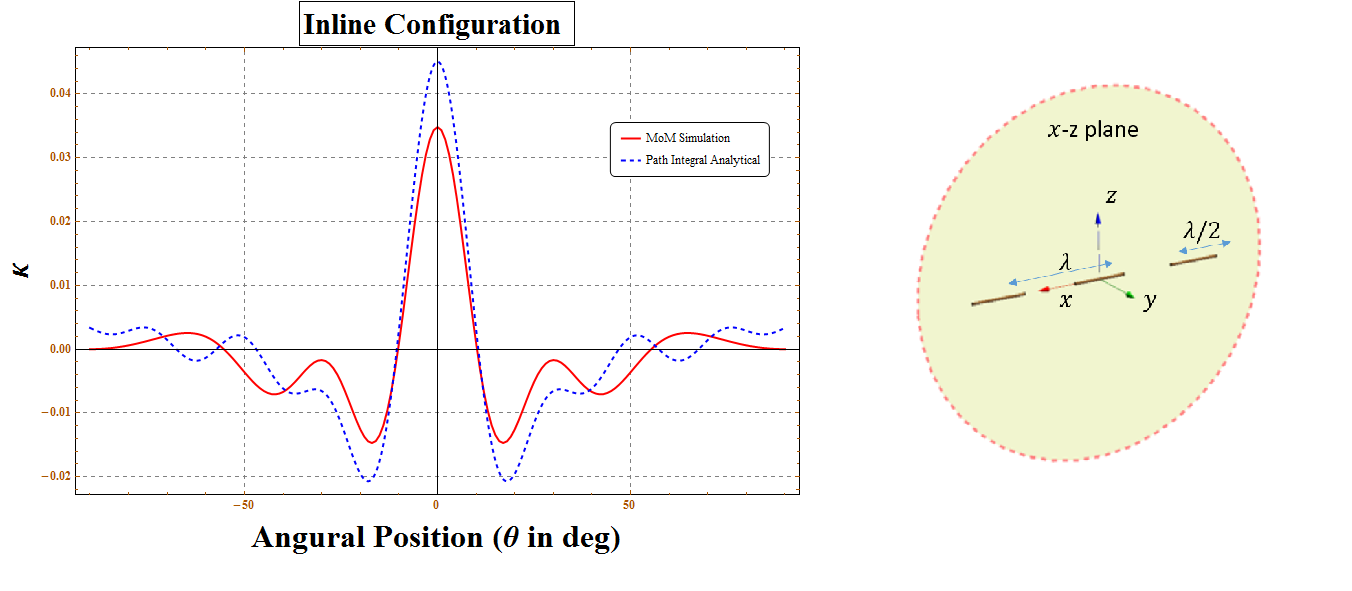}
\caption{The figure on the left shows $\kappa$ as a function of detector position. The red line indicates simulation done using Method of Moments. The blue dotted line is the result from the analytical formula based on path integral formalism derived in \cite{Scirep}. The figure on the right shows the dipole array configuration that was simulated.}
\end{figure*}

However, the distribution of $\kappa$ over 10 datasets is not always a normal distribution at all detector positions. For each $\kappa$ value, we have 15 sets of data corresponding to different slot combinations as well as background radiation value. Each data set has 3000 raw data points. If we plot a histogram of these 3000 points, for some of the combinations (around 2\% of the total number), the distribution is not normal as we would ideally like it to be, but slightly skewed. This means that sometimes, some combinations have some unwarranted fluctuations. This could have a simple cause like slight tilt in placing the slot in the slot stand which could have affected some of the data sets. As a result of this, at some detector positions, the 10 $\kappa$ values when plotted as a histogram also sometimes do not follow a normal distribution. This motivates our choice for the median instead of the mean as being more representative of our experiment. Mean is more prone to being affected by sudden fluctuations in numbers while median less so. That is why, we have chosen the median and its attendant inter quartile range as our inputs to the analysis instead of the mean and standard deviation. Even in the individual $\kappa$ estimations, we used the median of the 15 quantities which contribute to $\kappa$ instead of the mean for completeness. In this manuscript, we represent the data in a box plot, where we have the median of 10 $\kappa$ values as the representative value at each detector position and the inter quartile range plotted as the error bar.

\begin{itemize}
\item The median of the distribution
\item The interquartile range : The range that covers ($Q_1$) 25\% to ($Q_3$ )75\% of the data
\item Near outliers, Any data point beyond $Q1-1.5(Q3-Q1)$ or $Q3+1.5(Q3-Q1)$ is called a near outlier.
\item Far outliers, Any data point beyond $Q1-3(Q3-Q1)$ or $Q3+3(Q3-Q1)$ is called a far outlier.
\end{itemize}
In our results plot in the main text, we have chosen to not show the small number of outliers that were present in a few detector positions and shown the median and interquartile range as per standard convention.

\section*{Some calculations using parameters from radio astronomy}

Figure 12 shows the configuration that we have used to calculate $\kappa$ from parameters used in simulations of signals from the epoch of reionization of the early universe \cite{radiofour}.

We simulate the $\kappa$ as a function of detector position for an inline array of three dipole antennas. We consider a wire of radius $\lambda/100$ and length $\lambda/2$ with a centre fed port to be an array element \cite{radiofour} and measure $\kappa$ as a function of detector position at  $z = 10^3 \lambda$ which mimics far field. 
We have confirmed that the reciprocity theorem holds and the resultant graph for the three antennas acting as sources and acting as detectors is the same. (It is more common for antenna arrays to be used as detectors of signals from the early universe but easier for us to simulate the source based configuration).
For inline configuration, $\kappa$ computed from analytical formula \cite{Scirep} and numerical MoM method have similar modulation and magnitude.

$\kappa$ thus simulated from experimental astrophysical parameters and its convincing match with analytical path integral formula indicates that boundary condition effects are significant even in such experiments. The experimental astronomy community should take note of this result especially in the context of the current experiment which demonstrates that such order of magnitude for the deviation from superposition principle can be convincingly measured. Such effects will be especially significant in precision astronomy experiments where macroscopic error sources would have been eliminated. 

\section*{Effect of detector size on $\kappa$}

We have  simulated the effect on $\kappa$ from changing detector size inspired by observations in [15]. The detector size is varied as a fraction of the mean experimental aperture size value. It is varied uniformly for both E and H plane. Zero aperture size represents a screen detector. The uncertainty band has been formed by taking into account uncertainties in experimental parameters and the plot in Figure 13(a) indicates that our measured $\kappa$ value does not depend significantly on detector aperture size.

\begin{figure*}[t]
\begin{tabular}{cc}
\includegraphics[width=0.5\linewidth]{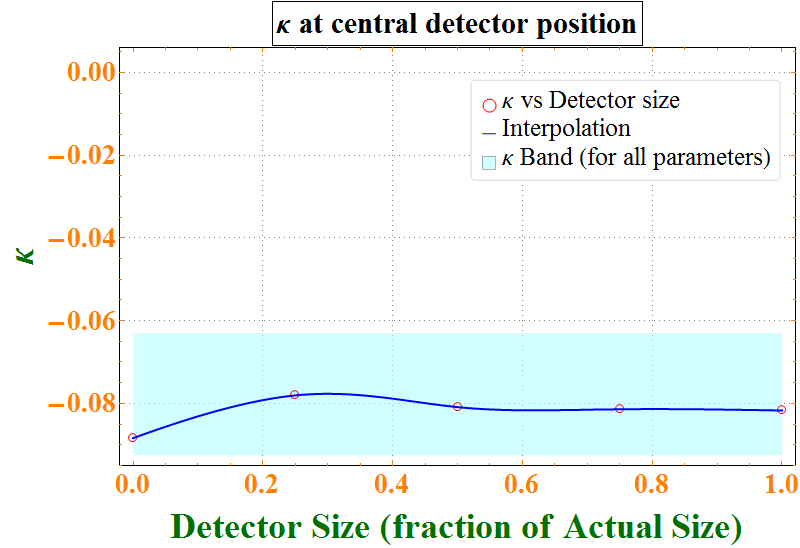}&\includegraphics[width=0.5\linewidth]{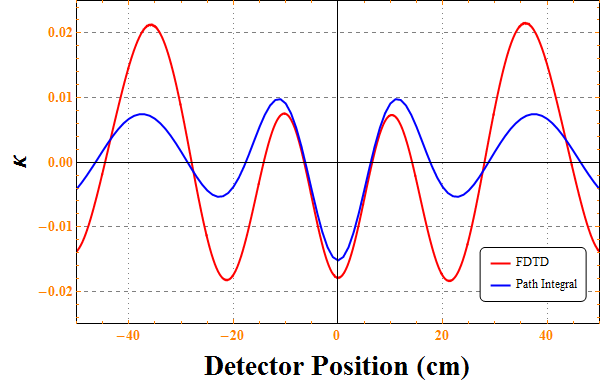}\\
(a) & (b)
\end{tabular}
\caption{(a) $\kappa$ at the central detector position as a function of varying detector aperture size. (b)The red line shows $\kappa$ as a function of detector position obtained using FDTD and parameters corresponding to figure 2b. The blue line is generated using path integral formalism. The effective slot width in path integral has been taken to be 7 cm as opposed to the actual width of 10 cm. As path integral does not capture material properties, it fails to capture the effect due to waves penetrating the material leading to an effective slot width which is smaller than actual one. This has been discussed in detail in [8]. FDTD on the other hand has material parameters as input so does not require the concept of effective slot width.}
\end{figure*}

 
\section*{Comparison of Path integral result with Finite Difference Time Domain simulations}

As earlier theory work on estimating the deviation from the superposition principle was done using path integral formalism [6,8] and FDTD simulations[9], we have also analysed our experiment using these techniques.
Figure 13(b) shows the $\kappa$ as a function of detector position for our slot experiment parameters using both path integral and FDTD.
The parameters used for the FDTD simulations were vertically polarized point dipole source, source to slot plane distance of 1m, slot plane to detector plane distance of 1m, simulation box = 2m along slot plane * 2.5m along the beam propagation direction, composite slots of 3.0mm aluminium sandwitched between 2.2.mm Eccosorb material, slot width of 10cm, inter-slot distance of 13cm. Eccossorb material model was for a paramagnetic material with permittivity of 11.107 and permeability of 1.912.
For the path integral simulations, we used slot width of 7cm, inter-slot distance of 13cm, wavelength of 5cm, point source with same source plane-slot plane and slot-plane-detector plane distances as FDTD.

While FDTD and Path Integral show reasonable match in magnitude and modulation, the plot does not match very well in magnitude with experiments as well as simulations based on Method of Moments (MOM). Both FDTD and Path integral have several shortcomings as compared to the full wave MOM based simulation. While these are 2D simulations, MOM is 3D. In MOM, one can define the horn source and horn detector while in both these methods, we use point source and point detector. Moreover, being 2D simulations, both FDTD and path integral assume an infinite slot height which is not the case in reality (slot height is 30 cm). Moreover, both path integral and FDTD will have errors as one moves away from the centre. Errors in path integral are due to finite integration domain whereas in FDTD, one has errors due to PML reflections.
Over and above these, path integral suffers from some additional limitations. The path integral formalism used in this paper as well in [6] and [8] is based on scalar field theory whereas FDTD can take into account source polarization. In path integral, material properties need to be accounted for by the concept of ``effective width". Also, we use the thin slot approximation whereas the slot actually has finite thickness. \\
Inspite of the above limitations, path integral formalism gives the same order of magnitude for $\kappa$ as FDTD as well as similar modulation as experiment. It also serves as a useful aid to distinguish between actual non-zeroness of $\kappa$ and error contributions as the contributions from the non classical paths can be turned off and on at will. It is thus a handy theoretical tool which is perhaps slightly too ideal to expect perfect match with experiments.

\section*{Acknowledgments}
GR and PU contributed equally to the work; GR, PU and SNS performed the experiment; SS helped in instrumentation and field based experience;  US conceived of and supervised the experiment; GR, PU, SNS and US contributed to modelling and analysis; US wrote the manuscript. We thank M. Arndt, K.B.Raghavendra Rao, A. Raghunathan, Barry Sanders, R. Subrahmanyan, N. Udayashankar and G.Weihs for useful discussions. We thank A. Sinha and D. Home for reading through the draft and helpful comments and discussions and A. Sinha also for theoretical insights. We thank Anjali P.S, Shreya Ray and Ashutosh Singh for their assistance in the preliminary stages of the project. We thank the staff members at the Gauribidanur Observatory for use of their facilities for the measurements, as well as mechanical workshop and carpentry section of RRI for their patience in building various versions of our equipment.

\clearpage
\onecolumngrid
\begin{center}
{\bf{Supplementary material} }
\end{center}
\vskip 2cm
\twocolumngrid 
\section*{Modification to the definition of the Sorkin parameter in the slot based interference experiments}

\renewcommand{\thefigure}{S\arabic{figure}}
\setcounter{figure}{0}
\begin{figure}
\centering
\includegraphics[scale=0.4]{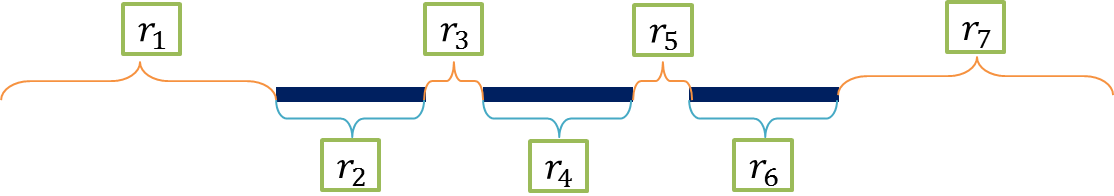}
\caption{The division of the slot plane into seven regions to explain the modification to the definition of the Sorkin parameter in slot based interference experiments }
\label{double}
\end{figure}

As seen in Figure S1, in case of a triple slot configuration, the slot plane can be divided into 7 regions. Let us denote the electric field emanating from region``x" as $\vec{E_x}$ and magnetic field as $\vec{H_x}$. Then naive application of the superposition principle gives us the following:\\
\begin{itemize}
\item $\vec{S}_{BG} \propto (\vec{E}_1 + \vec{E}_2 + \vec{E}_3 + \vec{E}_4 + \vec{E}_5 + \vec{E}_6 + \vec{E}_7)\times (\vec{H}_1 + \vec{H}_2 + \vec{H}_3 + \vec{H}_4 + \vec{H}_5 + \vec{H}_6 + \vec{H}_7)$  \\
\item $\vec{S}_{A}\propto (\vec{E}_1 + \vec{E}_3 + \vec{E}_4 + \vec{E}_5 + \vec{E}_6 + \vec{E}_7)\times (\vec{H}_1 + \vec{H}_3 + \vec{H}_4 + \vec{H}_5 + \vec{H}_6 + \vec{H}_7)$  \\
\item $\vec{S}_{B} \propto (\vec{E}_1 + \vec{E}_2 + \vec{E}_3 + \vec{E}_5 + \vec{E}_6 + \vec{E}_7)\times (\vec{H}_1 + \vec{H}_2 + \vec{H}_3 + \vec{H}_5 + \vec{H}_6 + \vec{H}_7)$  \\
\item $\vec{S}_{C} \propto (\vec{E}_1 + \vec{E}_2 + \vec{E}_3 + \vec{E}_4 + \vec{E}_5 + \vec{E}_7)\times (\vec{H}_1 + \vec{H}_2 + \vec{H}_3 + \vec{H}_4 + \vec{H}_5 + \vec{H}_7)$  \\
\item $\vec{S}_{AB} \propto (\vec{E}_1  + \vec{E}_3 + \vec{E}_5 + \vec{E}_6 + \vec{E}_7)\times (\vec{H}_1  + \vec{H}_3 + \vec{H}_5 + \vec{H}_6 + \vec{H}_7)$  \\
\item $\vec{S}_{BC} \propto (\vec{E}_1 + \vec{E}_2 + \vec{E}_3 + \vec{E}_5  + \vec{E}_7)\times (\vec{H}_1 + \vec{H}_2 + \vec{H}_3 + \vec{H}_5  + \vec{H}_7)$  \\
\item $\vec{S}_{CA} \propto (\vec{E}_1 + \vec{E}_3 + \vec{E}_4 + \vec{E}_5 + \vec{E}_7)\times (\vec{H}_1 + \vec{H}_3 + \vec{H}_4 + \vec{H}_5 + \vec{H}_7)$  \\
\item $\vec{S}_{ABC} \propto (\vec{E}_1 + \vec{E}_3 + \vec{E}_5 + \vec{E}_7)\times (\vec{H}_1 + \vec{H}_3 + \vec{H}_5 + \vec{H}_7)$  \\
\end{itemize}
We see that if we compute the numerator of $\kappa$ defined for slits
\begin{equation}
num(\kappa) = p_{A} + p_{B}+ p_{C} - p_{AB}-p_{BC}-p_{CA}+p_{ABC} \\= p_BG
\end{equation}
Here $\vec{S_\alpha}$ denotes the Poynting vector corresponding to a slot combination $\alpha$ and $p_\alpha$ denotes the real part of the Poynting vector which is the same as the magnitude of the Poynting vector in the radiative zone of the antenna.
Thus, we can define the new numerator to be 
\begin{equation}
p_{BG}- num(\kappa_{slits})
\end{equation}
which is zero when we apply the superposition principle incorrectly.

Thus, we define $\kappa$ for slots as
\begin{equation}
\kappa_{SLOT} = \frac{p_{BG}-(p_A+p_B+p_C- p_{AB}- p_{BC}-p_{CA}+p_{ABC})}{\max(p_{BG})} \\
\label{slotKappa}
\end{equation}
In our manuscript, $\kappa_{SLOT}$ will be denoted by $\kappa$.

\section*{Effect of back reflection from antenna/slots on measured values of $\kappa$}

When metal structures face each other and one is a source while the other a receiver, there is a possibility that the metal from the receiver will reflect some radiation back to the source as a result of which the radiation from the source will have an additional effect. In case of the current experiment which falls in the domain of precision experiments, it becomes very important to establish that such variations if any in the source radiation does not lead to a ``false" non-zero $\kappa$. In this experiment, there are three different ways in which such effects can play a role and we have systematically investigated all such effects and eliminated them as major contributors to our measured $\kappa$.\\
 The first investigation involved varying the source-receiver distance and measuring the receiver power as a function of detector position at increasing distances. While we found that there is some fluctuation and variation in the measured power for short distances between the source and receiver as expected, as the distance increased, the receiver power settled down to a constant value. This happens at about 50 cm distance between the flares of the two horn apertures. Our experiment has been carried out at a distance of 2 metres between the two horn flares and thus we are way beyond any distance where such effects matter. We also checked the same effects by removing the horn structure and just retaining the probe wires and found that the fluctuations again drop beyond a 50 cm distance between the horn flares. The presence of horn causes oscillations typical of a cavity structure which does not occur in the absence of horn. The presence of the horn of course is very useful in increasing the gain by orders of magnitude.  All oscillations and fluctuations occur at much closer horn to horn distance than in our experiment. We calculated the maximum effect that any such reflection can have on $\kappa$ and found that the effects are more than one order of magnitude lower than the measured $\kappa$. One could argue at this point that since increasing the distance between source and receiver gets rid of unwanted effects, why not make it even larger? We simulated the effects at a larger distance of 8 metres and while the simulated effects were lowered further, so did the actual power. In terms of measured power, in order to achieve more than three orders of precision, the minimum incident power should be in the micro-watt regime which is difficult to achieve at 8 metres distance given our maximum output from the signal generator. By choosing the distance to be 2 m instead of 8 m, the detected power roughly increases by a factor of 16 along with a larger predicted value of $\kappa$. This leads to a significant increase in the signal to noise ratio. This sealed our choice for 2 m distance between the horn flares which is 2.5 m distance between the probe wires for our experiment. \\
The second investigation requires us to use the concept of  $error~ \kappa$ as defined in the main text of the manuscript. \\
When we place slots directly in front of the horn source, the slots may reflect some fields back to the source which could also in principle cause the emitted source profile to change. As our experiment involves changing the slot configuration seven times, these back reflections can lead to changing source profiles for each such case which in turn can lead to a non-zero $\kappa$ simply due to the changing source term. We used the Method of Moments method to calculate the complex current in the probe wire for various slot combinations. Then we used these complex currents to appropriately scale the source wave function for each combination. As the current in the probe wire will change for different slot combinations due to the effect of back reflections, this will lead to different source powers for different combinations. These can then be used to estimate error $\kappa$. Figure S2 shows the error $\kappa$ calculated for our heterostructure material using a vertically oriented wire as a source. \\

\begin{figure}
\centering
\includegraphics[width=0.7\linewidth]{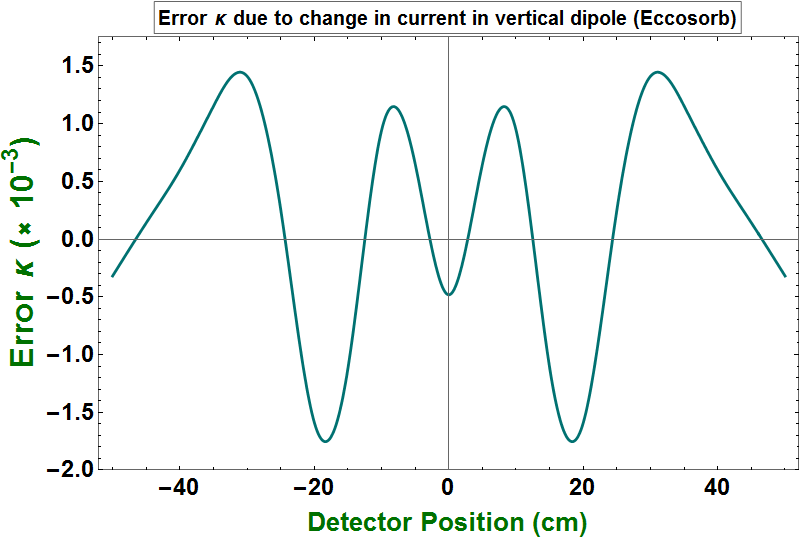}
\caption{$\kappa$ due to current change in dipole for heterostructure slots}
\end{figure}

We see that $\kappa$, if at all is due to change in source, is 0.0025 or less (almost 50 times less than the actual $\kappa$). \\
For metals, we take the horn source and carry current in the horn wire probe. We found the kappa computed from this method is five times less than simulated kappa(about 0.3).\\
The final investigation involved simulating error effects in absence of horn structure as well as metal probe. This will decide whether the simulated effects are due to reflections from the metal in the horn structure itself or mutual coupling between the probe wires. First, for the background combination, the complex electric and magnetic fields at the horn source aperture were calculated. The above fields are then imposed in the simulation i.e. they are assumed to be there without any probe wire as well as horn metal structure. The same source is imposed for every combination thus guaranteeing an invariant source. If there is a difference in the $\kappa$ graph between the case where the source is thus imposed to be invariant and when the actual horn structure with the probe wire acts as a source, then this can be attributed to spurious effects due to back reflection and mutual coupling etc. Figure S3 shows the $\kappa$ graph obtained for metal slots and screen detector with source-detector horn flare distance of 2 m. As can be seen, except a slight change at the centre, the two graphs show remarkable agreement thus ruling out back reflections as playing any dominant role in our measured values. One should note that since in the metal slot case, the effect can be ignored, the same will of course hold even for the heterostructure based experiments where the slots have been custom made to be maximally absorbant.

\begin{figure}
\centering
\includegraphics[width=0.7\linewidth]{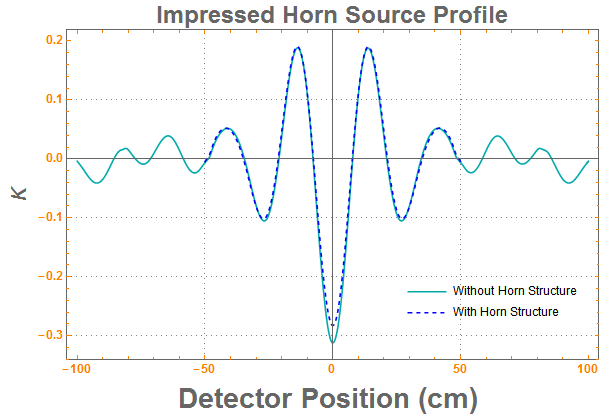}
\caption{Metal Slots, Screen detector at 100 cm from slot plane. Kappa with the imposed invariant source and the actual horn structure show remarkable agreement.}
\end{figure}

\section*{Movie file}
A movie file has been added to give a visual representation to some of the hugging paths that are mentioned in the text. As these effects are sub dominant, we have played with the colour scheme to ensure that the sub dominant contributions are visible. We have taken three slots (same material as experiment) and have the source on only from the left so that we can see the dominant paths move forward whereas some other much lower intensity ones make their way along the material and through a neighbouring slit. We have used the standard rainbow colour map to the lower range of power values in order to make them ``visible". We have also put a block in the forward direction to ensure that the forward moving dominant paths do not saturate the picture. This has been uploaded at http://www.rri.res.in/QuicLab/Kappa/

\section*{Data and analysis}
We have also uploaded the main experimental data as well as analysis files for $\kappa$ as a function of detector position at http://www.rri.res.in/QuicLab/Kappa/

\end{document}